\documentclass[revtex4]{emulateapj}

\bibpunct{}{}{;}{}{}{,}
\usepackage{times}
\usepackage{color}
\usepackage{appendix}
\usepackage{tabularx}
\usepackage{array}
\usepackage{graphicx}
\usepackage{epstopdf}
\usepackage{lscape}

\DeclareGraphicsExtensions{.eps}

\renewcommand{\theequation}{\thesection.\arabic{equation}}

\newcommand{\ssfr}{\hbox{$\psi_{\mathrm S}$}}
\newcommand{\tauv}{\hbox{$\hat{\tau}_{V}$}}
\newcommand{\mstar}{\hbox{$M_\star$}}
\newcommand{\msun}{\hbox{$M_\odot$}}
\newcommand{\aboh}{\hbox{$12+\log\textrm{(O/H)}$}}

\shorttitle{Timing the evolution of local galaxies}
\shortauthors{Pacifici et al.}

\begin{document}

\title{Timing the evolution of quiescent and star-forming local galaxies}

\author{Camilla~Pacifici,\altaffilmark{1,2} Sree Oh,\altaffilmark{3} Kyuseok Oh,\altaffilmark{4} Jaehyun Lee,\altaffilmark{5} and Sukyoung K. Yi.\altaffilmark{3}}

\altaffiltext{1}{NASA Postdoctoral Program Fellow, Goddard Space Flight Center, Greenbelt, MD 20771, USA}
\altaffiltext{2}{Yonsei University Observatory, Yonsei University, Seoul 120-749, Republic of Korea}
\altaffiltext{3}{Department of Astronomy, Yonsei University, Seoul 120-749, Republic of Korea}
\altaffiltext{4}{Institute for Astronomy, Department of Physics, ETH Zurich, Wolfgang-Pauli-Strasse 27, CH-8093 Zurich, Switzerland}
\altaffiltext{5}{Korea Astronomy and Space Science Institute, 776, Daedeokdae-ro, Yuseong-gu, Daejeon 305-348, Republic of Korea}

\begin{abstract}

Constraining the star formation histories (SFHs) of individual galaxies is crucial to understanding the mechanisms that regulate their evolution. Here, we combine multi-wavelength (ultraviolet, optical, and infrared) measurements of a very large sample of galaxies ($\sim$230,000) at $z<0.16$, with physically motivated models of galaxy spectral energy distributions to extract constraints on galaxy physical parameters (such as stellar mass and star formation rate) as well as individual SFHs. In particular, we set constraints on the timescales in which galaxies form a certain percentage of their total stellar mass (namely, 10, 50 and 90\%). The large statistics allows us to average such measurements over different populations of galaxies (quiescent and star-forming) and in narrow ranges of stellar mass. As in the downsizing scenario, we confirm that low-mass galaxies have more extended SFHs than high-mass galaxies. We also find that at the same observed stellar mass, galaxies that are now quiescent evolve more rapidly than galaxies that are currently still forming stars. This suggests that stellar mass is not the only driver of galaxy evolution, but plays along with other factors such as merger events and other environmental effects.

\end{abstract}


\section{Introduction}
\label{sec:intro}

Recent progresses in simulations of galaxy formation and the increasing quantity and quality of galaxy data are finally allowing us to assess the way galaxies form stars and evolve. The aims of such studies are numerous: to assess the effects of certain processes of galaxy formation (e.g., merger events, gas accretion, and feedback mechanisms) on the physical parameters and the morphology of galaxies; to link galaxies to their progenitors; and to understand how and when galaxies form, evolve in an active phase, and eventually stop forming stars to evolve passively.

The picture we have of galaxy evolution is a result of the combination of constraints obtained with different methodologies and datasets. A first approximation to the star formation history (SFH) of galaxies is the evolution of the star formation rate (SFR) density (\cite{madau1996}, \cite{schiminovich2005}, \cite{hopkins2006}, \cite{cucciati2012}, \cite{madau2014}) with redshift. These studies have unveiled the presence of a peak of star formation at $z\sim2$ and are good diagnostics for galaxy evolution models. The discovery of a bimodality in galaxy colors (\cite{blanton2003}, \cite{bell2004}, \cite{muzzin2013}, \cite{schawinski2014}, \cite{tomczak2014}, \cite{vulcani2014}) and the observation of this bimodality at all redshifts add constraints on the timescales of galaxy formation, especially regarding the processes that cause galaxies to transition from the \textit{blue cloud} (populated by star-forming galaxies) to the \textit{red sequence} (populated by quiescent galaxies) (\cite{schawinski2014}, \cite{smethurst2015}, \cite{taylor2015}). This transition of star-forming galaxies to quiescence is required also by the fact that the population of local quiescent galaxies is not consistent with a simple passive evolution of the population of quiescent galaxies at $z\sim1$ (\cite{schiavon2006}, \cite{gallazzi2014}). The SFH of the Universe can be studied not only observing galaxies at different redshifts, but also from the fossil record of present-day data (e.g., \cite{panter2003}, \cite{panter2007}, \cite{gonzalez2014}). Without strong prior assumptions on the shape of the SFHs of galaxies, evidence that SFHs are stochastic and non-parametric has been found by \cite{pacifici2013}, \cite{bauer2013}, and \cite{mcdermid2015}. This evidence is also an important piece of information in unveiling the mechanisms that are involved in the evolution of galaxies. Also, the relative importance of in-situ star formation and merger events has been studied by \cite{ownsworth2014} through the selection of the progenitors of local massive galaxies in the range $0.3<z<3$.

All these observations have triggered the development of more and more comprehensive cosmological simulations (e.g., \cite{springel2005}, \cite{croton2006}, \cite{somerville2008}, \cite{klypin2011}), including both dark matter and baryonic physics, along with the necessary prescriptions for stellar and AGN feedback. These simulations allow us to identify the progenitors of galaxies (\cite{moster2013}, \cite{behroozi2013}, \cite{papovich2015}) and study the mechanisms involved in the evolution of galaxies (e.g., minor and major mergers, stellar and AGN feedback, gas-accretion from filaments).

Galaxy observations allow us to observe the effects of such mechanisms at snapshots in time, while cosmological simulations can predict what mechanisms are responsible for the changes (in galaxy morphology, color, and physical parameters) between one snapshot and the following. In order to properly connect observations to simulations and learn about these mechanisms of evolution, we need to interpret large datasets of galaxies (spanning a large parameter space) with as-realistic-as-possible spectral-energy-distribution (SED) models.

With the approach described in \cite{pacifici2012}, we can compare cosmological simulations to real observations and derive meaningful constraints on the SFHs of galaxies. We build physically motivated SEDs of galaxies by combining the semi-analytic post processing of a large cosmological simulation with state-of-the-art models of the light from stars and gas and the attenuation by dust. We then use this large variety of SEDs to extract constraints from photometric and spectroscopic observations of galaxies. In this work, we focus on low-redshift galaxies selected from the Sloan Digital Sky Survey (SDSS). The combination of physically motivated SFHs and a large dataset gives us the opportunity to estimate the timescales of formation and SFH shapes of different populations of galaxies, covering a large stellar mass range for both quiescent and star-forming galaxies.

A comprehensive view on the SFHs of galaxies covering a broad range of physical parameters is very important to calibrate cosmological simulations and unveil the processes responsible for the evolution of galaxies. We start here a series of papers in which we constrain the SFHs of a large sample of galaxies at redshift $0.02<z<0.16$. This paper is organized as follows: in Section~\ref{sec:data} and ~\ref{sec:model} we present the multiwavelegth dataset and the modelling approach, respectively; in Section~\ref{sec:results}, we show and discuss the constraints we extract from the photometric fits; in Section~\ref{sec:concl}, we present a summary and our conclusions. In the following papers, we will compare our results with predictions from different cosmological simulations (Paper II), we will include spectroscopic constraints for a subsample of recently quenched galaxies (Paper III), we will select a subsample of galaxies and look for evidences of particular evolutionary mechanisms in the morphology of the galaxies (Paper IV), and we will assess the effects of environment (Paper V).

Throughout the Paper, we adopt a Chabrier initial mass function (IMF, \cite{chabrier2003}) and a standard $\Lambda$CDM cosmology with $\Omega_{\mathrm{M}}=0.3$, $\Omega_{\Lambda}=0.7$, $h=0.7$. Magnitudes are given in AB system.

\section{Data}
\label{sec:data}

\begin{figure*}
\begin{center}
\includegraphics[width=1\textwidth]{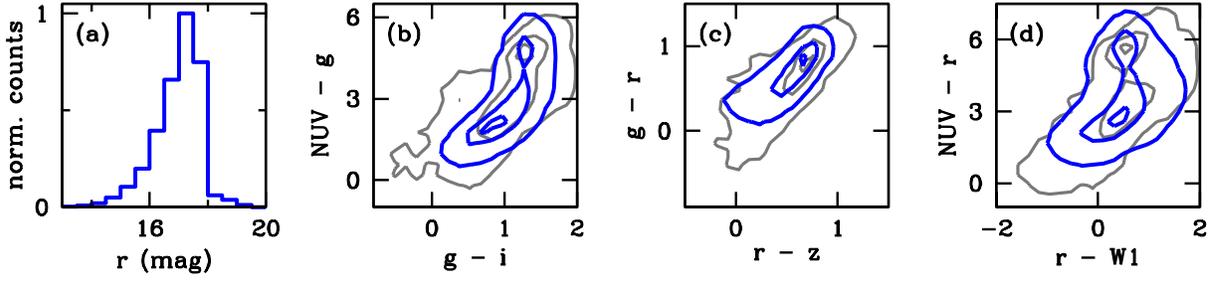}
\caption{Panel (a): $r$-band distribution of the galaxies in the sample. Panel (b): comparison between the observer-frame $NUV-g$ and $g-i$ colors of the galaxies in the sample (blue contours) and of the model galaxies in the spectral library (gray contours). Panel (c): same comparison for $g-r$ and $r-z$ colors. Panel (d): same comparison for $NUV-r$ and $r-W1$. The contours represent 1, 30 and 80\% of the maximum density. Our spectral library can accurately reproduce the distribution of the observed galaxies in color-color space.}
\label{fig:col}
\end{center}
\end{figure*}

We select a sample of galaxies from the Sloan Digital Sky Survey (SDSS) Data Release 10 (\cite{ahn2014}). SDSS provides us with optical photometry in five broad bands (\textit{ugriz}) and fiber spectroscopy in the wavelength range $3800<\lambda/\mathrm{\AA}<9200$ (and thus accurate spectroscopic redshifts) for about 1.8 million galaxies in the nearby Universe. We focus on the redshift range $0.02<z<0.16$ which includes the bulk of SDSS observations and covers the last 2 Gyr in the history of the Universe (512,961 galaxies)\footnote{We selected only galaxies with ``clean photometry'' flags as suggested from SDSS III DR10 tutorial.}. To include only galaxies with accurate photometry (we select Petrosian magnitudes), we apply the following criteria: $r<22.2$ mag, which is the magnitude limit for point sources (94\% of the galaxies are detected at $r<17.77$, which is the magnitude limit of the SDSS spectroscopic Main Sample; see Figure~\ref{fig:col}a); uncertainties in \textit{griz} bands smaller than 0.2 mag;\footnote{We do not include the $u$ band. See Section~\ref{sec:fitting} for more details.} similar photometric aperture in all bands (less than 3 kpc difference relative to the aperture in the $r$ band). This leaves us with 454,293 galaxies.

To better assess the young stellar component of the galaxies in the sample (\cite{salim2005}), we extend the coverage to ultraviolet wavelengths matching the sample with measurements from GALEX (Galaxy Evolution Explorer, General Release 6/7; \cite{martin2005}). GALEX samples near-ultraviolet (NUV; $\lambda=2316$ {\AA}) and far-ultraviolet (FUV; $\lambda=1539$ {\AA}) wavelengths. We match the 454,293 galaxies with GALEX NUV detections (matching radius of 3 arcsec) that are within 0.5 degrees from each tile center and have uncertainties smaller than 0.5 magnitudes. After the matching, we are left with 233,816 galaxies, 58\% of which are also detected in the FUV band.

In order to constrain also the oldest stellar populations, we need to include measurements at near-infrared wavelengths. The Two Micron All Sky Survey (2MASS, \cite{skrutskie2006}) provides us with measurements in $J$, $H$, and $K$ bands, but unfortunately only for the most massive galaxies (about half of the sample). This is not preferable, because we wish to keep good number statistics for low-mass galaxies and have comparable constraints (and thus comparable error bars) for all galaxies. We decide thus to move to slightly larger wavelengths and match the 233,816 galaxies in the sample with the Wide-field Infrared Survey Explorer (WISE; \cite{wright2010}). WISE is a full-sky survey in the 3.4, 4.6, 12 and 22 $\micron$ mid-infrared bandpasses. For this study, we consider only the 3.4 $\micron$ observations (W1 hereafter) since fluxes at longer wavelengths are contaminated by the emission by warm dust which is not included in our spectral modelling approach (see Section~\ref{sec:model}). Adopting again a matching radius of 3 arcsec and following the guidelines of \cite{chang2015}, we find that 232,343 galaxies out of 233,816 are detected in the W1 band. W1 photometric measurements are systematically fainter than measurements of the same objects using deeper and higher spatial resolution data from the Spitzer Space Telescope. This bias occurs because background levels were overestimated in the source extraction process. We thus apply the magnitude correction calculated by \cite{chang2015} to enhance the flux in W1 as a function of the effective radii of the galaxies. The median correction is 0.25 mag and it is applied whenever galaxies are larger than 0.5 arcsec.

We use the emission-line catalogue by \cite{oh2011} (OSSY)\footnote{The OSSY catalog is based on SDSS DR7 spectra. Emission-line measurements are available for 92\% of the galaxies in the DR10 sample used in this work.} to identify active galactic nuclei (AGNs) among the sources in the sample. We find that 449 galaxies show broad emission lines and are thus classified as `type-1' AGNs (\cite{oh2015}). In type-1 AGNs, the optical continuum is dominated by non-thermal emission, thus we remove these objects from the sample (we are left with 231,894 galaxies). Using the standard \cite{baldwin1981} line-diagnostic diagram and the conservative criterion of \cite{kauffmann2003a}, 38\% of the galaxies (89,077) are identified as `type-2' AGNs (20,241 out of these 89,077 are classified as Seyfert galaxies). According to \cite{kauffmann2003a}, the type-2 AGN continuum emission does not affect the estimates of the physical parameters derived from optical photometric fits. At longer wavelengths, \cite{assef2010} provide an AGN color selection using WISE observations. Using their criterion ($W1-W2>0.85$ in Vega magnitudes), only about 1\% of the galaxies in our sample seem to show AGN features in the broad bands, suggesting that at these low redshifts the stellar light is not strongly contaminated by AGN light. We thus decide to neglect possible contaminations by type-2 AGNs when running the photometric fits.

All magnitudes are corrected for foreground galactic extinction with the dust map of \cite{schlafly2011} assuming a galactic extinction law with $R_V=3.1$ (\cite{fitzpatrick1999}). We do not correct W1 as the correction is negligible.

\section{Modeling approach}
\label{sec:model}

To derive physical parameters and SFHs of the 231,894 galaxies in the sample from multi-band photometry, we need to appeal to a comprehensive spectral modelling technique. Reliable constraints can be obtained only when the adopted library of model SEDs can properly reproduce the observed data and includes a large variety of SFH shapes. We adopt the approach described in \cite{pacifici2012} which includes realistic SFHs from cosmological simulations, stellar and nebular emission computed consistently, and a comprehensive prescription to model the attenuation by dust. We introduce here the library of model SEDs and the fitting technique.

\subsection{Library of model spectral energy distributions}
\label{sec:library}

We assemble a model spectral library including physically motivated star formation and chemical enrichment histories derived from a post-treatment of the Millennium cosmological simulation (\cite{springel2005}) using the semi-analytic models of \cite{delucia2007}. This approach allows us to explore a wide range of (non-parametric) SFHs, including declining, rising, roughly constant, bursty and smooth evolutionary shapes, in addition to realistic chemical enrichment histories. We generate 1.5 million galaxy SFHs, selecting randomly the redshift of observation in the range $0.01<z<0.17$ (appropriate to cover the range of the sample described in Section~\ref{sec:data}) and the evolutionary stage at observation between 0 and 6 Gyr in lookback time (see Sections 2.1 and 3.1.2 of \cite{pacifici2012}). In other words, this resample allows the model galaxies to be younger than what the Millennium simulation predicts at a given redshift of observation. To further widen the range of physical properties probed by this library, we re-sample the `current' (i.e., averaged over a period of 10Myr before a galaxy is looked at) star formation properties of each star-forming model galaxy (about 60\% of the galaxies in the library), while leaving unchanged the properties of quiescent galaxies.\footnote{We define as `star-forming' the galaxies with specific SFR (SSFR, the SFR divided by the stellar mass; $\ssfr=\psi/M_{\ast}$) larger than 0.01 Gyr$^{-1}$ (\cite{kimm2009}).} In practice, for star-forming model galaxies, we redraw the SSFR randomly in the range $-2<log(\ssfr/\mathrm{Gyr}^{-1})<1.5$ and the current gas-phase oxygen abundance in the range $7<\aboh<9.4$.

We combine this set of SFHs with the latest version of the \cite{bruzual2003} stellar population synthesis models, reprocessed with the photoionization code {\small CLOUDY} (\cite{ferland1996}) as in \cite{charlot2001} to compute the nebular emission consistently with the emission by stellar populations of different ages (neglecting the contribution by stars older than 10 Myr, which produce negligible ionizing radiation). We include dust attenuation using a more comprehensive implementation of the \cite{charlot2000} two-component dust model. In this work, the slope of the attenuation curve is fixed in the birth clouds ($n=-1.3$), and it is drawn randomly in the range $-1.0<n<-0.4$ in the diffuse interstellar medium (ISM) to reflect uncertainties about the spatial distribution of dust and the orientation of a galaxy (see \cite{pacifici2012}; \cite{chevallard2013}). We take the total effective optical depth of the dust $\tauv$ to be randomly distributed between 0 and 4 for star-forming galaxies and between 0 and 2 for quiescent galaxies to account for galaxies with two main populations (e.g., a star-forming disk and a quiescent bulge).

In Figure~\ref{fig:col}, we show a comparison between the observer-frame colors of the galaxies in the sample (blue contours) and the colors predicted by the spectral library (gray contours) for the entire redshift range. The spectral library can well reproduce the observed colors of the galaxies in the sample. $NUV-g$ (Figure~\ref{fig:col}b) and $NUV-r$ (Figure~\ref{fig:col}d) colors clearly show the bimodality in the galaxies: quiescent galaxies are characterized by red colors (large $NUV-g$ and $NUV-r$), while star-forming galaxies lie in the \textit{blue} region (low $NUV-g$ and $NUV-r$). Since we are requiring a detection in $NUV$, we might be missing a population of very red galaxies that fall below the sensitivity of GALEX. In the sample presented in Section~\ref{sec:data}, about 60\% of the galaxies are \textit{blue} in $g-r$ color ($g-r<0.75$). If we do not ask for a detection in $NUV$ (454,293 galaxies, see Section~\ref{sec:data}), the optical colors spanned by the galaxies are similar to those of the sample presented in Section~\ref{sec:data} (as shown in Figure~\ref{fig:col}c), and still about 60\% of the galaxies are \textit{blue} according to the same color criteria. The sample presented is thus not biased by the request of GALEX NUV detection. In Figure~\ref{fig:col}(b) and (d), a small fraction of the observed galaxies (about 2\%) falls beyond the edge of the model library at large $NUV-g$ and $NUV-r$ colors. These outliers are all characterized by large uncertainties in the NUV measurements ($\simeq 0.5$ mag) and thus can still be safely compared to the model spectral library.

\subsection{Fitting procedure}
\label{sec:fitting}

\begin{figure}
\begin{center}
\includegraphics[width=0.47\textwidth]{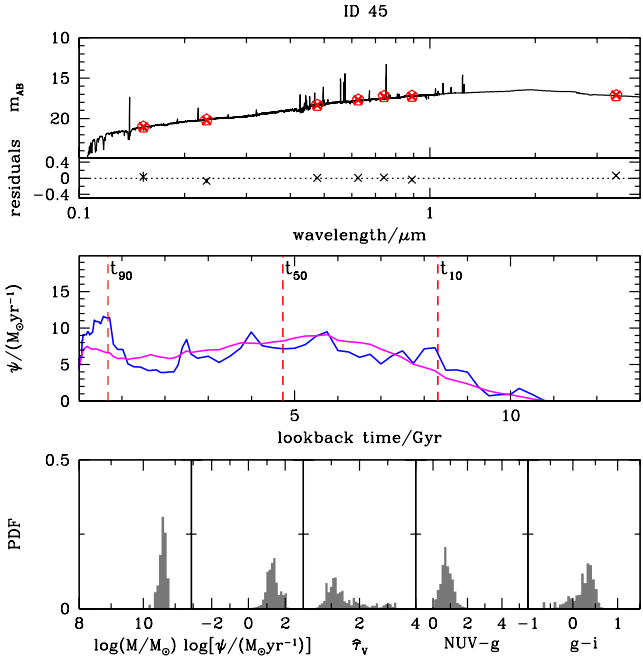}
\caption{A fit to an example galaxy from the sample. Top panel: observed photometric magnitudes (red crosses), best-fit photometric magnitudes (red open diamonds), best-fit SED in full resolution (black solid line), and residuals between the observed and best-fit magnitudes (black crosses). Middle panel: likelihood-weighted average SFH derived from the first 10 best-fit models (blue solid line) and using all models in the library (magenta solid line); lookback times at which the galaxy reaches 10\% ($t_{10}$), 50\% ($t_{50}$), and 90\% ($t_{90}$) of the total stellar mass formed (red vertical dashed lines). Bottom panels: probability density function of (from left to right) stellar mass, SFR, optical depth of the dust, and rest-frame dust-corrected $NUV-g$ and $g-i$ colors.}
\label{fig:fit}
\end{center}
\end{figure}

\begin{figure}
\begin{center}
\includegraphics[width=0.47\textwidth]{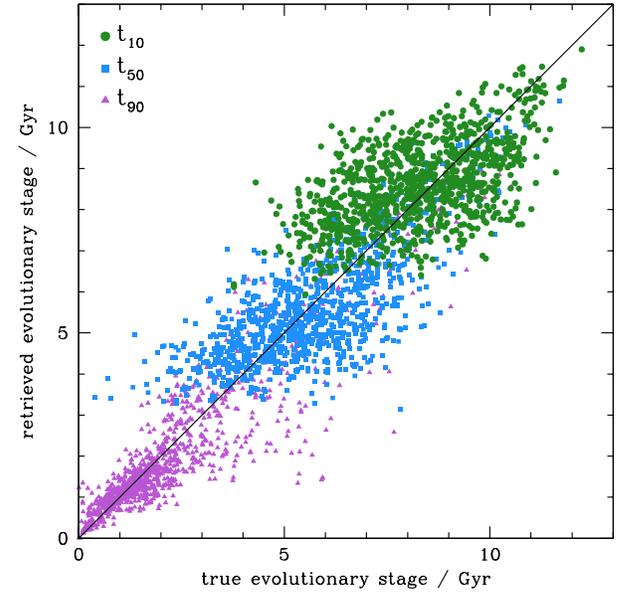}
\caption{Simulation of retrievability of the $t_{10}$ (green circles), $t_{50}$ (blue squares), and $t_{90}$ (purple triangles) evolutionary stages (lookback times at which the galaxies form 10\%, 50\%, and 90\% of the total stellar mass, respectively). We pick 1000 random galaxies from the sample and extract the evolutionary stages from the best-fit SFHs derived from the photometric fits. We re-fit these 1000 model galaxies and retrieve the evolutionary stages from the best-estimate SFHs as described in Section~\ref{sec:fitting}. We then compare the obtained evolutionary stages with the true values. This procedure allows us to quantify the accuracy and uncertainty to which the evolutionary stages are constrained. See text for details.}
\label{fig:ttsimu}
\end{center}
\end{figure}

We use a Bayesian approach as in \cite{pacifici2012} to compare the SEDs of each of the 231,894 galaxies in the sample to the SEDs of the galaxies in the model spectral library. In practice, for each galaxy in the sample, we calculate the likelihood of each of the 1.5 million model galaxies (at redshifts $z=z_{observation}\pm0.01$) to reproduce the observed SED and we build probability density functions (PDFs) of: stellar mass ($\mstar/\msun$), star formation rate [$\psi$/(\msun yr$^{-1}$)], total effective optical depth of the dust ($\tauv$), and rest-frame dust-corrected $NUV-g$ and $g-i$ colors. We also extract the best-estimate SFHs by averaging the first 10 best-fit model SFHs weighted by their likelihood.

In the likelihood fit, we include the 7 photometric bands ($FUV, NUV, g, r, i, z, W1$) presented in Section~\ref{sec:data}.\footnote{We apply a minimum uncertainty of 5\% on all bands to account for the fact that the uncertainties might be sometimes underestimated and systematic errors might have been ignored.} We do not include the $u$ band because we find the observed flux systematically fainter (0.1 magnitudes on average) than what is predicted by the model library.\footnote{The exclusion of this photometric measurement is not critical as the ultraviolet light is already probed by the GALEX $NUV$ band, for which the comparison with the model library does not show any systematic bias.} Although this deviation is not large, we find that if we do not include the $u$ band in the fits, the distributions of the residuals between observed and best-fit magnitudes in neighboring bands ($NUV$, $g$, and $r$) is Gaussian, while if we include the $u$ band, these distributions become asymmetric, to correct for the discrepancy between models and observations in the $u$ band. This discrepancy could be caused by complications in the data reduction or by uncertainties in the stellar evolution models (see for example \cite{worthey1996}). Another possible explanation for the discrepancy could be an inaccurate match between GALEX and SDSS photometry. Given the large point spread function of GALEX, source confusion could generate issues when matching the catalogs (see for example \cite{moustakas2013}). However, if the mismatch was severe for a significant number of galaxies, we would expect a discrepancy much larger than 0.1 magnitudes in the $u$ band. To check for this effect, we limit the sample to the SDSS galaxies with a GALEX match within 2 arcsec instead of 3 arcsec (87\% of the galaxies), with no neighboring galaxies within this radius (62\% of the galaxies), and with exposure time in $NUV$ greater than 1000 s (34\% of the galaxies). The discrepancy in the $u$ band for this subsample of galaxies is the same as for the full sample, suggesting that this is not the cause of the discrepancy. In Section~\ref{sec:results}, we will check that our results do not change when using this more conservative subsample. We note that the uncertainties in the photometry are included in the fits, thus in case of short observations or possible contaminations GALEX uncertainties will be larger than the average and thus GALEX measurements will be down-weighted in the fits.

To check that the correction we apply on WISE W1 magnitudes does not introduce any significant bias, we compare the predicted magnitudes from the best-fit models in $J$, $H$, and $K$ bands with the observed ones from 2MASS (after matching to the SDSS $r$-band position with a 2 arcsec matching radius). We remind the reader that only massive galaxies are detected by 2MASS. Best-fit magnitudes are slightly brighter than the observed ones in all three bands, but the difference is on average within 1$\sigma$. We thus are positive that the matching and magnitude corrections are not introducing severe biases.

In Figure~\ref{fig:fit}, we show a fit to an example galaxy extracted from the sample. The top panel shows the observed broad-band magnitudes in the 7 filters (red crosses), the best-fit model magnitudes (red open diamonds), the best-fit SED in full resolution (black solid line), and the residuals between the observed and best-fit magnitudes (black crosses). The middle panel of Figure~\ref{fig:fit} shows the likelihood-weighted average SFH derived from the first 10 best-fit models in blue ($t=0$ corresponds to the lookback time at the redshift of observation) and from the full model library in magenta. The SFH extracted using the whole library is naturally smoother than the one averaged on 10 models. The red vertical dashed lines represent the lookback times at which the galaxy forms 10\% ($t_{10}$), 50\% ($t_{50}$), and 90\% ($t_{90}$) of the total stellar mass formed in the entire lifetime. From now on, we refer to $t_{10}$, $t_{50}$, and $t_{90}$ as evolutionary stages. In the bottom panels of Figure~\ref{fig:fit}, we show the PDFs of stellar mass, SFR, optical depth of the dust, and rest-frame dust-corrected $NUV-g$ and $g-i$ colors. The typical uncertainties on these five quantities are 0.1 dex, 0.5 dex, 0.5, 0.4 mag, and 0.1 mag respectively.

To derive the best-estimate SFHs we made the choice of averaging over the first 10 best-fit models instead of using all the models in the library, as it is done for the single parameters (e.g., stellar mass, SFR, etc.). This choice allows the procedure to be computationally much less expensive given the number of galaxies in the sample, still providing a good representation of the SFHs derived including in the weighted average all models with likelihood greater than zero. This is the case because the PDFs of the extracted parameters are generally well behaved, i.e. they are single-peaked. To verify this, we extract PDFs of the light-weighted age, which is the parameter most sensitive to the SFH shape. The uncertainties on the light-weighted age are smaller than 0.2 dex for 96\% of the sample, thus most likely the PDFs are not double-peaked. In the Appendix, we show the fits of few example galaxies for which we have computed the average SFHs using the first 10 best-fit models and the full model library. The shapes of the SFHs compare well with one other.

To estimate the uncertainties on the derived evolutionary stages ($t_{10}$, $t_{50}$, and $t_{90}$), we extract best estimates of these quantities fitting a sample of model galaxies and we compare the results with the true values. In practice: 1) we extract from the fits the best-fit photometry and SFHs of 1000 galaxies randomly picked from the observed sample; 2) for each of the 1000 galaxies, we fit the best-fit photometry (adopting the photometric uncertainties as in the original observed sample and removing from the model library the best-fit model) and we extract the best-estimate SFH; 3) we compare the evolutionary stages of the best-estimate SFHs of the 1000 galaxies with the true values from the fitted models. In Figure~\ref{fig:ttsimu}, we compare the extracted evolutionary stages with the true values for the 1000 model galaxies. We find that $t_{10}$ (green circles) is mildly overestimated by a median of $0.38$ Gyr and the median uncertainty is 0.93 Gyr. The mild overestimate of $t_{10}$ can be explained by the fact that the luminosity of a small fraction of stars at such old ages can be hidden by the luminosity of more recent stars. The fits account for this possibly hidden mass and thus return a slightly older $t_{10}$ compared to the true value. $t_{50}$ (blue squares) is well recovered (median accuracy of $-0.02$ Gyr) with median uncertainty of $0.77$ Gyr. $t_{90}$ (purple triangles) is similarly well recovered with median accuracies and uncertainties of $-0.07$ and 0.32 Gyr, respectively.

The sample down to $r<22.2$ is complete to $\log(M_{\ast}/M_{\odot})= 9.5$ including the whole redshift range. For the spectroscopic main sample ($r<17.77$), the mass completeness limits are $\log(M_{\ast}/M_{\odot})=8.5, 9.0, 9.7, 10.4, 10.9$ at $z=0.03, 0.05, 0.09, 0.12, 0.16$, respectively.

\section{The star formation histories of local galaxies}
\label{sec:results}

Constraints on the SFHs of the galaxies in the sample (Section~\ref{sec:data}) can show us whether different types of galaxies (e.g., currently quiescent or star-forming) evolve following different paths. In this Section, we first explore the relation between SFH evolutionary stages and stellar mass, being the latter a driver of galaxy evolution.\footnote{Because of the mass completeness limits of the sample, at low stellar masses, the results will be dominated by galaxies observed at $z<0.05$, while at high stellar mass, observations in the full redshift range will contribute.} We then assess the rest-frame dust-corrected colors of the galaxies in the sample as a function of SFH properties. A careful assessment of the uncertainties is needed to draw meaningful conclusions.

\subsection{Evolutionary stages of observed galaxies}
\label{sec:evostage}

\begin{figure*}
\begin{center}
\includegraphics[width=1\textwidth]{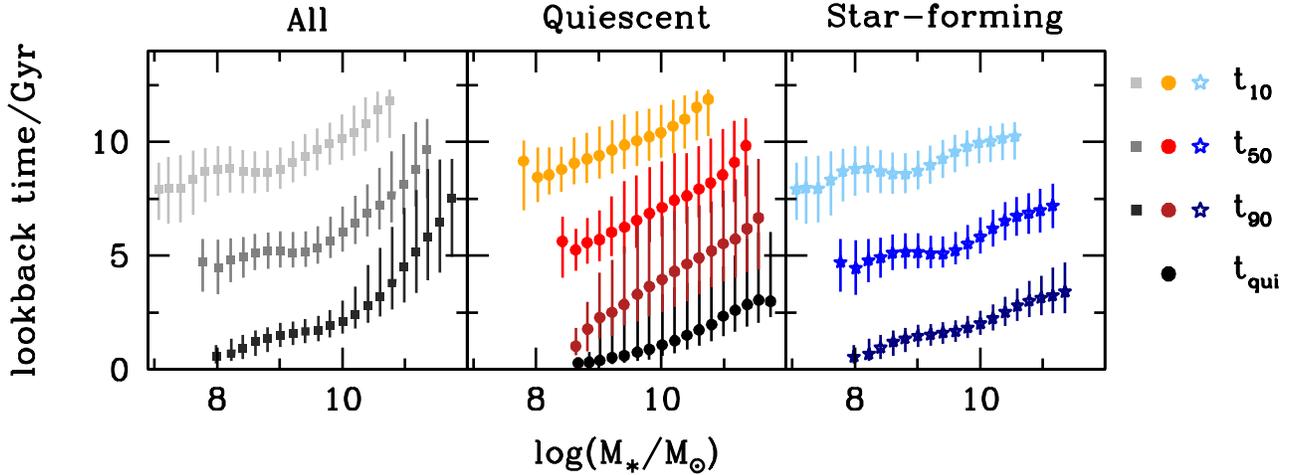}
\caption{Evolutionary stages derived for all (left-hand-side panel), quiescent (central panel), and star-forming (right-hand-side panel) galaxies in the sample. The evolutionary stages are calculated from the best-estimate SFHs. Symbols represent the median values of (from light to dark colours) $t_{10}$, $t_{50}$, and $t_{90}$. The time of quiescence ($t_{qui}$) is shown only for quiescent galaxies (central panel, black circles). In all panels, the downsizing effect is prominent, i.e., low-mass galaxies are younger than high-mass galaxies.}
\label{fig:tsteps}
\end{center}
\end{figure*}

\begin{figure*}
\begin{center}
\includegraphics[width=1\textwidth]{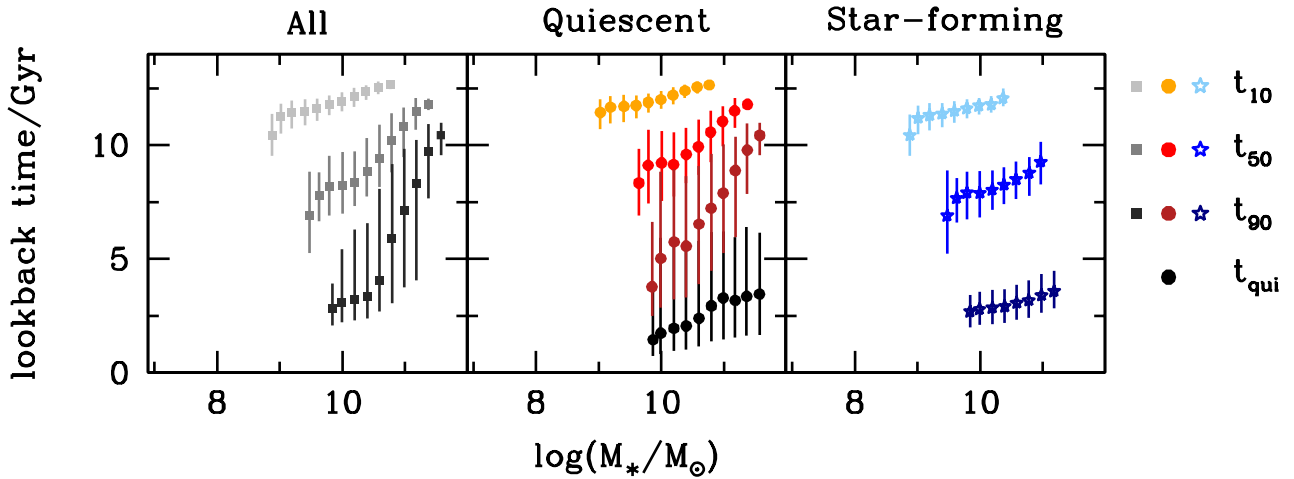}
\caption{Evolutionary stages derived from the model SFHs of \cite{delucia2007}. The tracks include all progenitors. The format is the same as the one in Figure~\ref{fig:tsteps}.}
\label{fig:tstepsmod}
\end{center}
\end{figure*}

Using the procedure described in Section~\ref{sec:fitting}, we derive best estimates of stellar mass, SFR, SFH, and $NUV-g$ and $g-i$ colors for each galaxy in the observed sample. For each SFH, we measure the lookback times at which the galaxy reaches 10\% ($t_{10}$), 50\% ($t_{50}$), and 90\% ($t_{90}$) of the total stellar mass formed and the actual stellar mass at these evolutionary stages (i.e., accounting for the fraction of stellar mass returned to the ISM by an evolving stellar population). For quiescent galaxies, we also measure the lookback time at which each galaxy reaches quiescence [$\log(\ssfr/\mathrm{Gyr}^{-1})=-2$] and remains quiescent for the rest of its lifetime ($t_{qui}$). Being the galaxies observed in the redshift range $0.02<z<0.16$, we correct all evolutionary stages for the age of the Universe at the redshift of observation (i.e., for all galaxies, we set $t=0$ at $z=0$).

We show in Figure~\ref{fig:tsteps} the median evolutionary stages of all (left-hand-side panel, squares), quiescent (central panel, circles, 34\% of the total sample), and star-forming (right-hand-side panel, stars, 66\% of the total sample) galaxies in bins of stellar mass ($0.2$~dex wide). The error bars show the 16-to-84 percentiles of the distributions per bin. In each panel, the colors indicate, from light to dark, $t_{10}$, $t_{50}$, $t_{90}$, and $t_{qui}$ (the latter only in the case of quiescent galaxies). In all panels, we see the effect of downsizing, i.e., high-mass galaxies form earlier than low-mass galaxies. When including all galaxies (left-hand-side panel), there appear to be a transition mass where the slope of the evolutionary stage as a function of stellar mass changes. These smooth transitions happen at $\sim 10^9$ (at $t_{10}$), $\sim 10^{9.5}$ (at $t_{50}$), and $\sim 10^{10.2}$~\msun (at $t_{90}$). A similar transition mass ($3\times10^{10}$~\msun) is suggested by \cite{kauffmann2003b} to set the transition between young low-surface-mass-density low-concentration galaxies (low stellar mass) and old galaxies with high surface mass density and high concentration (high stellar mass).

Quiescent galaxies (central panel in Figure~\ref{fig:tsteps}) evolve rapidly. For example, those detected with $10^{10.5}$~\msun at observation form the bulk of the stellar mass (10\% to 90\% of the total stellar mass) between on average $9.3\pm1.1$ (when the stellar mass is $\sim 10^{9.6}$~\msun) and $4.2\pm2.0$ Gyr (when the stellar mass is close to the value at observation). Also, the relation between the evolutionary stage and mass is steep and do not show a strong transition as it appears when all galaxies are included (left-hand-side panel). The transition might nonetheless be hidden in the dispersion of the measured evolutionary stages, which is large for quiescent galaxies (possibly because quiescent galaxies are likely merger remnants holding complex SFHs). This suggests that the possible time at which galaxies shut off their star formation covers a large range in lookback time at all masses. Also, we would expect to find median quiescence times ($t_\mathrm{qui}$) older than 3 Gyr for high-mass galaxies (e.g., \cite{williams2009}, \cite{brammer2011}, \cite{whitaker2013}). In this analysis, we are not considering that the transitory SSFR between star-forming, blue galaxies and quiescent, red-sequence galaxies evolve with redshift. According to \cite{fumagalli2014} (see also \cite{whitaker2012} and \cite{karim2011}), the SFR of star forming and quiescent galaxies increases by about 10 times between $z=0.5$ ($\sim5$ Gyr) and 2 ($\sim10.2$ Gyr) at all stellar masses. Thus, the transitory SSFR increases by the same amount. For simplicity, we do not include this variation in the calculation of $t_\mathrm{qui}$ and we keep the transitory SSFR fixed at $\psi_S=0.01$ Gyr$^{-1}$. This is the reason why we do not observe large $t_\mathrm{qui}$ at the highest masses. We will take this detail into account in the next paper, when we will compare the results of this work with the predictions from different cosmological simulations. It is interesting to note a small population of low-mass quiescent galaxies (central panel of Figure~\ref{fig:tsteps}, $\mstar<10^9\msun$, 226 galaxies) that are most likely post-starburst galaxies. To confirm their nature, an analysis of the spectroscopic features of these galaxies would be required. We will investigate this in Paper III.

\cite{mcdermid2015} find similar results to ours using a different technique to analyse the spectra of 260 nearby early-type galaxies from the ATLAS$^{\mathrm{3D}}$ survey. Similarly to what we find in this work, they confirm that low-mass galaxies have more extended SFHs and as a result are younger than more massive galaxies. The most massive galaxies in our sample (described in Section~\ref{sec:data}) form half of the total stellar mass in about 4--5 Gyr, while, according to \cite{mcdermid2015}, galaxies with stellar masses larger than $10^{11.0}$\msun (assuming a correction of $\sim0.3$ dex from dynamical to stellar masses; \cite{taylor2010}) form faster and reach 50\% of their stellar mass in less than 2 Gyr. This discrepancy could be attributed to the differences in the selection of the samples (different criteria to select quiescent and early-type galaxies) and also to the large dispersion we observe in the estimates of $t_{50}$ at large stellar masses. With respect to low-mass galaxies, we find our results in agreement with the estimates by \cite{mcdermid2015}, i.e. that galaxies with stellar masses between $10^{9}$ and $10^{9.5}$\msun form half of their stellar mass in about 8 Gyr.

Star-forming galaxies (right-hand-side panel in Figure~\ref{fig:tsteps}) evolve slower compared to quiescent galaxies forming the bulk of their stellar mass between on average $8.3\pm0.8$ (when the stellar mass is $\sim 10^{9.6}$~\msun) and $1.5\pm 0.7$ Gyr (when the stellar mass is very close to the stellar mass at observation). The dispersions of the measured evolutionary stages are smaller than in the case of quiescent galaxies. At $10^{9}$ (at $t_{10}$), $10^{9.5}$ (at $t_{50}$), and $10^{10}$\msun (at $t_{90}$), we can observe mild changes in the slopes of the evolutionary stages, suggesting a difference in the SFHs of low-mass and high-mass star-forming galaxies (\cite{pacifici2013}). This change in slope is larger when considering all galaxies (left-hand-side panel in Figure~\ref{fig:tsteps}). This is caused by the relative proportion of quiescent and star-forming galaxies in different stellar mass bins. At low stellar masses ($\sim 10^{9.5}$\msun) 88\% of the galaxies are classified as star-forming, while at higher stellar masses ($\sim 10^{11}$\msun) this fraction drops to 40\%.

With this analysis, we quantify the galaxy evolutionary stages and we find that galaxies with the same stellar mass can evolve in different ways (e.g., the evolutionary stages of quiescent and star-forming galaxies of the same stellar mass happen at different lookback times). Environment and morphology can be among the causes for the different evolutionary paths. We will investigate these options in Papers IV and V.

The results presented here have been obtained without including measurements in $J$, $H$, and $K$ photometric bands because these are available only for the brightest galaxies. Before proceeding, we need to quantify how much information we are missing by not fitting to these near-infrared bands. We thus measure the evolutionary stages of the 109,238 galaxies for which 2MASS magnitudes are available by fitting to 10 photometric bands ($FUV$, $NUV$, $g$, $r$, $i$, $z$, $J$, $H$, $K$, and $W1$) and we compare them with the measurements obtained without including 2MASS. We find that for galaxies with $\log(M_{\ast}/M_{\odot})<10.8$ the two results compare well with each other. Above that stellar mass, star-forming galaxies still compare well, while quiescent galaxies reach 50\% and 90\% of their stellar mass about 3 Gyr earlier than what is measured without including 2MASS magnitudes in the fits. This discrepancy affects about 15\% of the galaxies in the sample (massive, quiescent galaxies). Reasons for this difference might be related to the actual need of 2MASS observations to properly constrain the SFHs of the most massive galaxies, or to the way 2MASS photometry is extracted for the reddest galaxies (2MASS apertures are in same cases twice as large as SDSS apertures), or a combination of the two. A deeper analysis is not crucial for the present paper, but we will investigate it further in the next work.

The evolutionary stages we here measure from observations can be compared to the predictions by cosmological simulations to assess whether such simulations can reproduce the observed quantities, and also to identify the mechanisms that cause the variations in the timescales at different stellar masses and star-formation activity. As an example, we calculate the same evolutionary stages as in Figure~\ref{fig:tsteps} from the original SFHs predicted by the semi-analytic model of \cite{delucia2007} (Figure~\ref{fig:tstepsmod}). In practice, we ``observe'' the model galaxies at random redshifts between 0.02 and 0.16. We build model SEDs by applying the same stellar, gas and dust prescriptions as explained in Section~\ref{sec:library}. We then apply the same selection criteria and instrumental limits as we did for the observed sample (see Section~\ref{sec:data}).\footnote{To properly represent the bulk of the observed sample, we apply a magnitude cut in the $NUV$ band at 24 mag.} Finally, we can derive the evolutionary timescales ($t_{10}$, $t_{50}$, and $t_{90}$) for the selected model galaxies (plotted in Figure~\ref{fig:tstepsmod}). We find that the relation between the evolutionary stages and stellar mass is similar to what we derive from the data, i.e., massive galaxies are older than low-mass galaxies. Also, we find that the difference in evolution between star-forming and quiescent galaxies is similar to what is observed: quiescent galaxies (central panel in Figure~\ref{fig:tstepsmod}) evolve faster than star-forming galaxies (right-hand-side panel in Figure~\ref{fig:tstepsmod}). The differences we find between the model predictions and what we observe from the data appear to be in the absolute values of the evolutionary stages. Model galaxies tend to form faster than what we observe at all masses. Adjustments in some parameters of the models (e.g., the strength of feedback mechanisms and the star-formation efficiency) could reduce these differences. We note that the use of different cosmological parameters in the simulations (e.g., using the \textit{WMAP} or \textit{Planck} results) does not lead to any notable difference in the model SFHs (private communication with Aldo Rodr\'iguez-Puebla in reference to \cite{rodriguez2016}) and thus cannot explain the difference between the measured and predicted evolutionary timescales. We do not go into the details of the comparison here. In Paper II, we will explore the causes of the difference between the observational data and the model predictions and extend the comparison to other cosmological simulations.

\subsection{Tracks}

\begin{figure}
\begin{center}
\includegraphics[width=0.47\textwidth]{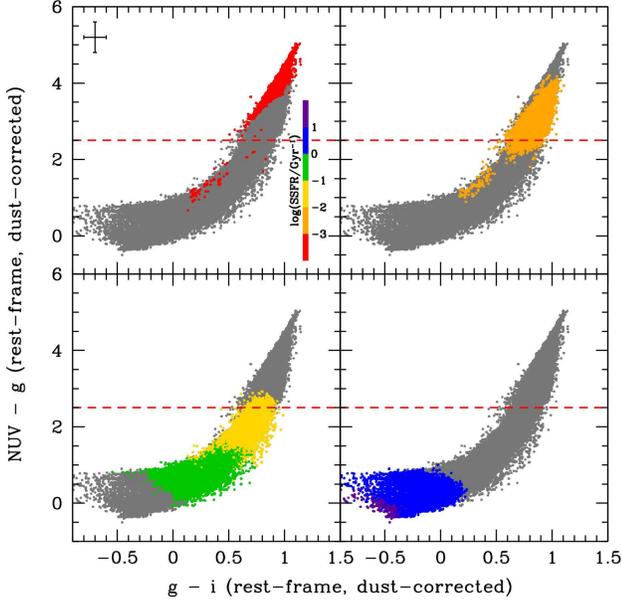}
\caption{Best estimates of $NUV-g$ and $g-i$ colors of the galaxies in the sample. In each panel we show the full sample in grey and a subsample according to cuts in specific SFR. The shade (shown in the top-left panel), from red to blue, represents the increasing of the specific SFR. Quiescent galaxies $[\log(\ssfr/\mathrm{Gyr}^{-1})]<-2$ show $NUV-g$ colors larger than 2.5 (red dashed horizontal line). The typical uncertainty on the derived colors is shown in the top-left panel.}
\label{fig:trackssfr}
\end{center}
\end{figure}

\begin{figure}
\begin{center}
\includegraphics[width=0.47\textwidth]{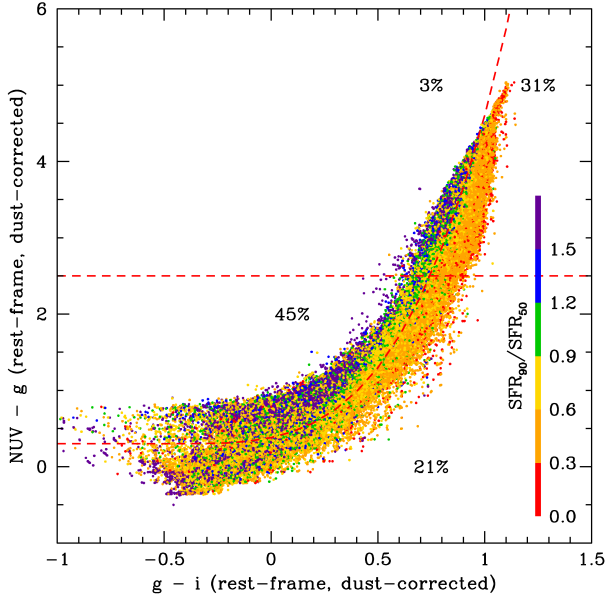}
\caption{Best estimates of $NUV-g$ and $g-i$ colors of the galaxies in the sample. The shade (shown on the right-hand side), represents the ratio between the SFRs at $t_{90}$ (SFR$_{90}$) and at $t_{50}$ (SFR$_{50}$). An empirical demarcation line (curved, red, dashed line; see text for details) is defined to distinguish galaxies with small and large SFR$_{90}$/SFR$_{50}$. The horizontal red dashed line is the same as the one in Figure~\ref{fig:trackssfr}. The percentages of galaxies in each region are also shown.}
\label{fig:tracksfr}
\end{center}
\end{figure}

\begin{figure}
\begin{center}
\includegraphics[width=0.47\textwidth]{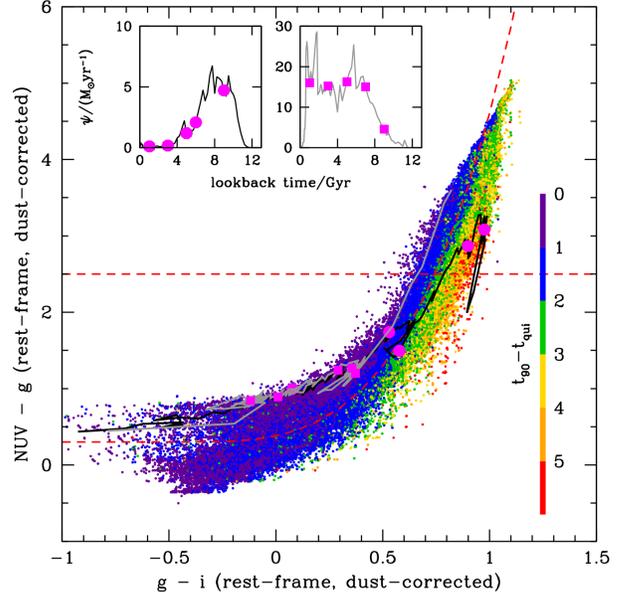}
\caption{Best estimates of $NUV-g$ and $g-i$ colors of the galaxies in the sample. The shade (shown on the right-hand side), represents, from blue to red, increasing difference between $t_{90}$ and $t_{qui}$. The demarcation lines (red dashed lines) are the same as those in Figures~\ref{fig:trackssfr} and \ref{fig:tracksfr}. The best-estimate SFHs of two example galaxies are shown in the small panels (black and gray). The relative color-color histories are shown in the big panel. Big square and circle magenta points mark $t=$1,3,5,6, and 9 Gyr on both the color-color and star-formation histories.}
\label{fig:trackdt}
\end{center}
\end{figure}

In Section~\ref{sec:evostage}, we have shown that the galaxies in the observed sample (described in Section~\ref{sec:data}) experience different histories of formation depending on the stellar mass at the time of observation. Low-mass galaxies evolve slower than high-mass galaxies. Such differences in the SFHs generate differences in the colors of the galaxies at the time of observation. We here explore the best-estimate rest-frame dust-corrected UV-optical colors of the galaxies in the sample as a function of SFH characteristics, as extracted from the fits (see Section~\ref{sec:fitting}).

Figure~\ref{fig:trackssfr} shows the $NUV-g$ versus $g-i$ rest-frame dust-corrected colors of all the galaxies in the sample (in grey in all four panels). The shades represent the current best-estimate specific SFR, increasing from red to blue. The border between quiescent and star-forming galaxies [$\log(\ssfr/\mathrm{Gyr^{-1}})=-2$] corresponds to a $NUV-g$ color of about 2.5 (red horizontal dashed line). We thus confirm that the $NUV-g$ color is a good tracer of specific SFR. Less than 1\% of the galaxies in the sample show $NUV-g<2.5$ and specific SFRs indicating quiescence. These are all low-mass galaxies for which the constraints look poor. Degeneracies among the parameters (age, SFR, dust attenuation, and metallicity) are probably causing the fit to choose extreme solutions.

In Figure~\ref{fig:tracksfr}, we show the same color-color measurements as in Figure~\ref{fig:trackssfr} with the shades representing the \textit{shape} of the SFHs. To quantify the shape of each best-estimate SFH, we calculate the ratio between the average SFR at $t_{90}$ (SFR$_{90}$) and the average SFR at $t_{50}$ (SFR$_{50}$). The SFRs are averaged over 0.6 Gyr around each evolutionary-stage value, i.e., $t\pm0.3$ Gyr. If the ratio SFR$_{90}$/SFR$_{50}$ is larger than 1, the SFH is likely rising as a function of time. If instead the ratio is smaller than 1, the SFH is likely a declining function of time. This is of course an approximation of the behavior of galaxies along their histories, given that galaxies do not necessarily evolve smoothly following rising or declining functions. The presence of strong bursts of star formation can mimic SFR$_{90}$/SFR$_{50}$ ratios that are not fully representative of a smooth rising or declining evolution. However, with the large number statistics we have, we observe trends in SFR$_{90}$/SFR$_{50}$ ratio in the color-color plane. This suggests that despite galaxies are not necessarily characterized by smooth rising or declining functions, the shapes of the SFHs can be reasonably quantified by the SFR$_{90}$/SFR$_{50}$ ratio. Figure~\ref{fig:tracksfr} shows that galaxies with similar SFR$_{90}$/SFR$_{50}$ are characterised by similar colors. We define a purely empirical demarcation line between the two regimes as $NUV-g=[0.5+(g-i)]^{3.6} +0.3$ (red curved dashed line): galaxies described by rising SFHs tend to occupy the region above the line (48\% of all galaxies), while those described by declining SFHs appears to be located below the line (52\% of all galaxies). It is interesting to note that 3\% of the galaxies in the sample are classified as quiescent, but show a rising SFH. These are most likely post starburst galaxies that experience a very rapid quenching of the star formation. These objects are extremely interesting and will be studied in detail in Paper III. In particular, we will include spectroscopic constraints such as Balmer absorption and D4000 measurements to estimate the age of the last burst of star formation with higher precision.

In Figure~\ref{fig:trackdt}, we show again the same color-color values as in the previous two figures. The shades here represent the time difference between $t_{90}$ (i.e., the lookback time at which a galaxy reaches 90\% of the total stellar mass formed) and $t_{\mathrm{qui}}$ (i.e., the lookback time at which a galaxy, that is not forming stars at observation, reaches quiescence). In the case of star-forming galaxies, $t_{\mathrm{qui}}$ is always 0, as the galaxies have not reached quiescence yet. Green to red shades represent long $t_{90}-t_{\mathrm{qui}}$, i.e., the declining of star formation is happening slowly. Blue and purple shades represent short $t_{90}-t_{\mathrm{qui}}$, i.e., the quenching of star formation is a fast process. The red dashed lines are the same as in Figure~\ref{fig:tracksfr}. It is important to note that this time difference cannot be interpreted as a \textit{quenching time}, because galaxies with a declining SFH (as in Figure~\ref{fig:tracksfr}, where SFR$_{90}$/SFR$_{50}$$< 1$) start switching off their star formation before reaching $t_{90}$, thus $t_{90}-t_{\mathrm{qui}}$ is only a fraction of the real quenching time. Nevertheless, we find that the difference between $t_{90}$ and $t_{\mathrm{qui}}$ (which we measure accurately) can already help distinguish galaxies on the $NUV-g$ versus $g-i$ diagram. Long time differences are observed for most of the quiescent galaxies. The quiescent galaxies above the red dashed lines (3\% of the sample) show instead very short time differences, supporting the idea that these galaxies might be post-starburst galaxies.

In the small panels of Figure~\ref{fig:trackdt}, we show the best-estimate SFHs of two example galaxies from the sample observed in the quiescent region of the plot ($NUV-g>2.5$), and on separate sides of the empirical curve introduced in Figure~\ref{fig:tracksfr} (red dashed curve). The evolution of the two tracks in color-color space are represented on top of the small colored dots. We observe that the black SFH is bell-shaped and fairly smooth (left-hand side small panel). This galaxy slowly become quiescent between lookback time $t=5$ Gyr and $t=3$ Gyr. Before observation, the black track moves briefly back to the star-forming region because of a quick event of star formation and moves then back to quiescence by the time of observation. The gray SFH (right-hand side small panel) is more bursty than the black one. It evolves as a rising function of time and drops fast in the last Gyr. The gray track in the color-color diagram is very similar to the black track for most of the history and moves on a different path only in the last Gyr to end in the top (quiescent) left-hand (short $t_{90} - t_{\mathrm{qui}}$) side of the digram.

From the evidences shown in Figure~\ref{fig:trackdt}, we could conclude that a UV-optical color-color diagram (\cite{yi2005}, \cite{williams2009},\cite{arnouts2013}) can help distinguish between galaxies that quench rapidly their star formation, like the gray track, and galaxies that slowly transition from star forming to quiescent, like the black track (as suggested also by \cite{schawinski2014} where quiescent galaxies are divided between early-type fast-quencher and late-type slow-quencher using morphological indicators). We have to point out however that the uncertainties on the rest-frame dust-corrected colors (shown in Figure~\ref{fig:trackssfr}) are of the order of or larger than the difference between the tracks (0.2-0.3 dex). Such uncertainties must be taken into account when assessing the tracks of galaxies on a rest-frame dust-corrected color-color diagram. More observations (especially spectroscopic observations) can help reduce the uncertainties and allow us to draw better conclusions on the path galaxies take to move from the blue cloud to the red sequence.

\section{Summary and conclusion}
\label{sec:concl}

Assessing the characteristics of the SFHs of individual galaxies is key to understanding the processes that are responsible for the formation and evolution of galaxies (major and minor mergers, gas accretion, feedback processes). The combination of photometric (and spectroscopic) observations and realistic, physically motivated models of the spectral energy distributions of galaxies allows us to shed some light onto such processes and quantify their relative importance along the histories of  galaxies.

In this Paper, we use our comprehensive spectral library of galaxies (including physically motivated SFHs from cosmological simulations and state-of-the-art models of the light from stars and gas and the attenuation by dust) to interpret photometric observations (GALEX, SDSS, and WISE) of 231,894 galaxies at redshifts $0.02<z<0.16$. For each galaxy, we derive best-estimate, realistic SFHs and we analyze their characteristics in terms of \textit{shape} and evolutionary stages. In particular, we derive the lookback times at which each galaxy reaches 10, 50, and 90\% of the total stellar mass formed. The uncertainties on these measurements are $\sim 1$, 0.8, and 0.3 Gyr, respectively. 

We find the following.
\begin{itemize}
\item Low-mass galaxies appear to have more extended SFHs than high-mass galaxies (in agreement with what we see also in simulations, e.g., \cite{delucia2007}, \cite{lee2013}).
\item When considering all galaxies, there appear to be a transition mass at $\sim 10^{10}$~\msun, above which the slope of the evolutionary stages as a function of mass gets steeper. This change in slope is milder when considering only quiescent or only star-forming galaxies, thus it is in part caused by the relative proportion of quiescent and star-forming galaxies in different stellar mass bins.
\item Galaxies observed as quiescent evolve rapidly and (for $\sim 10^{10.5}$~\msun at observation) form the bulk of the stellar mass (10\% to 90\% of the total stellar mass) between on average $9.3 \pm 1.1$ and $4.2 \pm 2.0$ Gyr.
\item Galaxies observed as star-forming (at the same stellar mass) evolve slower compared to quiescent galaxies and form the bulk of their stellar mass between on average $8.3 \pm 0.8$ and $1.5 \pm 0.7$ Gyr.
\item On a UV-optical color-color diagram (rest-frame, dust-corrected magnitudes), galaxies with different SFH shapes occupy different regions. Quiescent galaxies are located above $NUV-g > 2.5$. Galaxies characterized by rising SFHs (i.e., the SFR keeps growing from formation to observation) are located above the empirical function $NUV-g=[0.5+(g-i)]^{3.6}+0.3$ that roughly divides the color-color sequence in half. We find a small population (3\% of the observed sample) of post-starburst galaxies that are quiescent, but show rising SFHs.
\item Galaxies that shut off the star formation slowly or fast follow different tracks on a UV-optical color-color diagram. The differences between such tracks are however of the order of the uncertainties on the rest-frame dust-corrected colors, thus we need spectroscopic observations to reduce the uncertainties before being able to draw strong conclusions.
\end{itemize}

This Paper opens doors to many further studies of the SFHs of individual galaxies. Firstly, we can compare the derived evolutionary stages with predictions from different cosmological simulations. Such comparison will help calibrate the parameters of cosmological simulations (e.g., feedback and star-formation efficiency, relative importance of mergers and gas accretion) and help generate more and more realistic mock catalogs of galaxies. Such mock catalogs can then be used to simulate observations and best exploit current and future telescopes. Secondly, we can explore in more detail the features of post-starburst candidates by appealing to spectroscopic observations. Thirdly, we can assess the characteristics of SFHs as a function of morphology. We can identify merger events in the SFHs and look for correlations with observed weak and strong merger features. Finally, we can investigate the relations between timescales of formation and environment to constrain what cause certain galaxies to reach quiescence faster than others. We will address these questions in the following papers of the series.

\acknowledgments

We thank the Referee for the nice and useful report that helped us improve the paper. We thank Barbara Catinella, St\'{e}phane Charlot, Luca Cortese, Anna Feltre, Eric Gawiser, Michaela Hirschmann and Rory Smith for useful discussions. CP acknowledges the KASI-Yonsei Joint Research Program for the Frontiers of Astronomy and Space Science funded by the Korea Astronomy and Space Science Institute and support by an appointment to the NASA Postdoctoral Program at the Goddard Space Flight Center, administered by USRA through a contract with NASA. SKY acted as the head of the research group and the corresponding author and acknowledges support from the National Research Foundation of Korea (Doyak 2014003730). Funding for SDSS-III has been provided by the Alfred P. Sloan Foundation, the Participating Institutions, the National Science Foundation, and the U.S. Department of Energy Office of Science. The SDSS-III web site is http://www.sdss3.org/. SDSS-III is managed by the Astrophysical Research Consortium for the Participating Institutions of the SDSS-III Collaboration including the University of Arizona, the Brazilian Participation Group, Brookhaven National Laboratory, Carnegie Mellon University, University of Florida, the French Participation Group, the German Participation Group, Harvard University, the Instituto de Astrofisica de Canarias, the Michigan State/Notre Dame/JINA Participation Group, Johns Hopkins University, Lawrence Berkeley National Laboratory, Max Planck Institute for Astrophysics, Max Planck Institute for Extraterrestrial Physics, New Mexico State University, New York University, Ohio State University, Pennsylvania State University, University of Portsmouth, Princeton University, the Spanish Participation Group, University of Tokyo, University of Utah, Vanderbilt University, University of Virginia, University of Washington, and Yale University. This publication makes use of data products from the Wide-field Infrared Survey Explorer, which is a joint project of the University of California, Los Angeles, and the Jet Propulsion Laboratory/California Institute of Technology, funded by the National Aeronautics and Space Administration. GALEX (Galaxy Evolution Explorer) is a NASA Small Explorer, launched in 2003 April. We gratefully acknowledge NASAÕs support for construction, operation, and science analysis for the GALEX mission, developed in cooperation with the Centre National d'Etudes Spatiales of France and the Korean Ministry of Science and Technology.

\def\aj{AJ}
\def\araa{ARA\&A}
\def\apj{ApJ}
\def\apjl{ApJ}
\def\apjs{ApJS}
\def\apss{Ap\&SS}
\def\aap{A\&A}
\def\aapr{A\&A~Rev.}
\def\aaps{A\&AS}
\def\mnras{MNRAS}
\def\pasp{PASP}
\def\pasj{PASJ}
\def\qjras{QJRAS}
\def\nat{Nature}
\def\aplett{Astrophys.~Lett.}
\def\aas{AAS}
\let\astap=\aap
\let\apjlett=\apjl
\let\apjsupp=\apjs
\let\applopt=\ao

\section*{Appendix}
\renewcommand\thefigure{A\arabic{figure}}
\setcounter{figure}{0}
\renewcommand\thetable{A\arabic{table}}
\setcounter{table}{0}

We present the fits to a few example galaxies for which we compute the average SFHs using the first 10 best-fit models and all models with likelihood greater than zero. We find that the shapes of the two SFHs compare well with one another. The average obtained with all models is smoother simply because the bursts are more and more washed out when including more and more models in the average. The galaxies with large error bars in the UV measurements show some discrepancy between the two derived SFHs, highlighting the importance of the UV data to solve the degeneracies among age, SFR, and dust attenuation.

\begin{figure*}
\begin{center}
\includegraphics[width=0.32\textwidth]{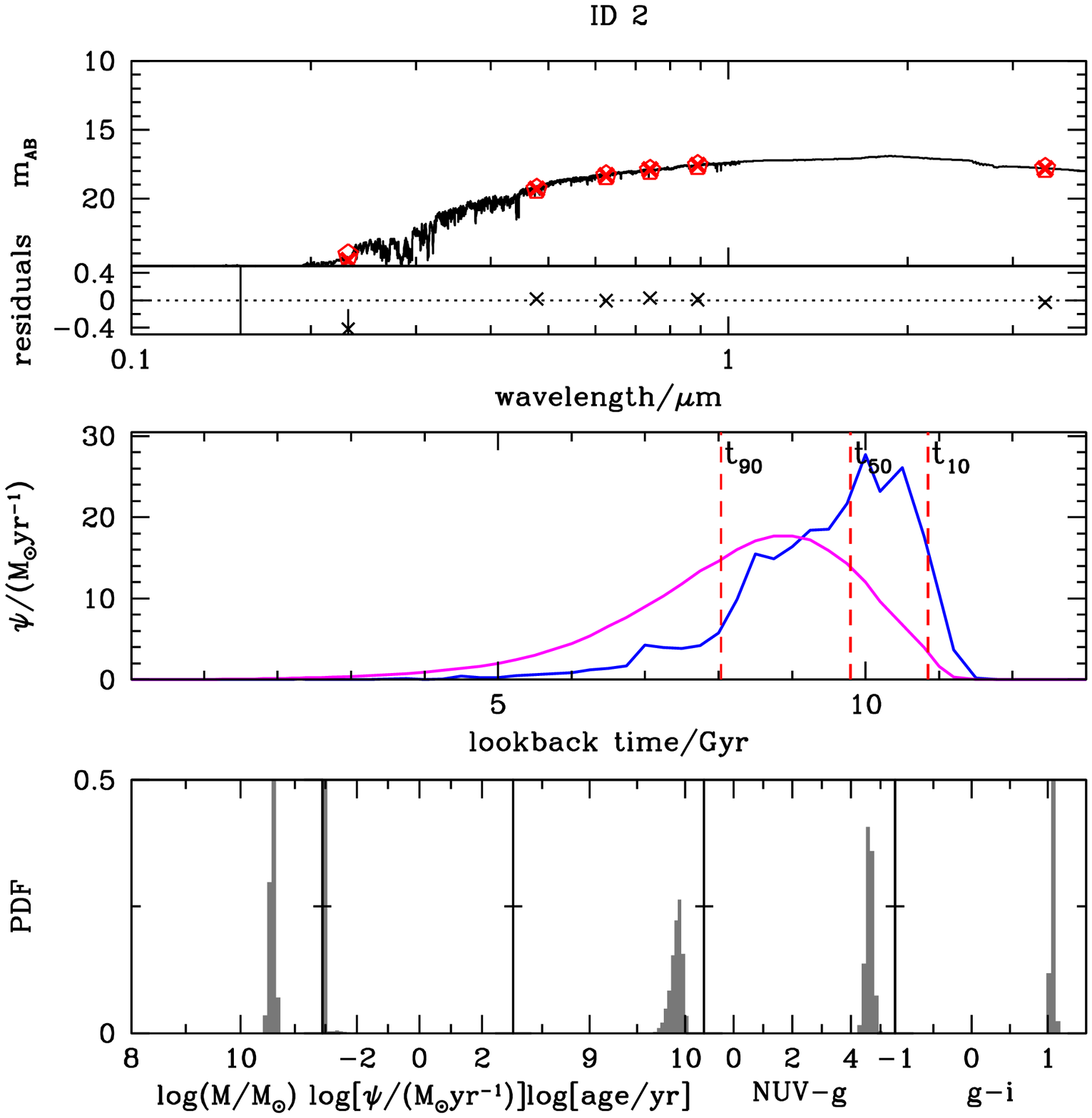}
\includegraphics[width=0.32\textwidth]{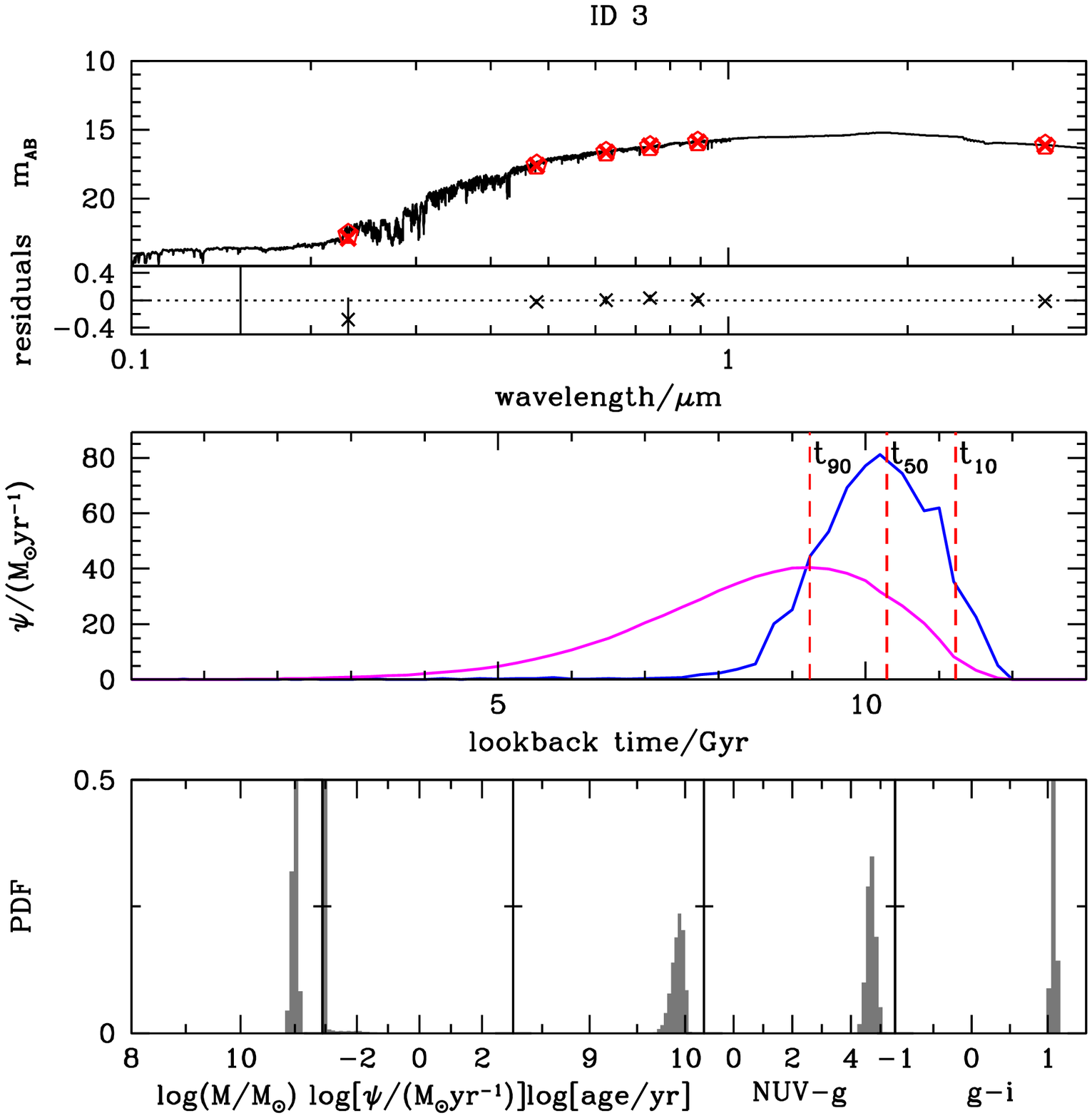}
\includegraphics[width=0.32\textwidth]{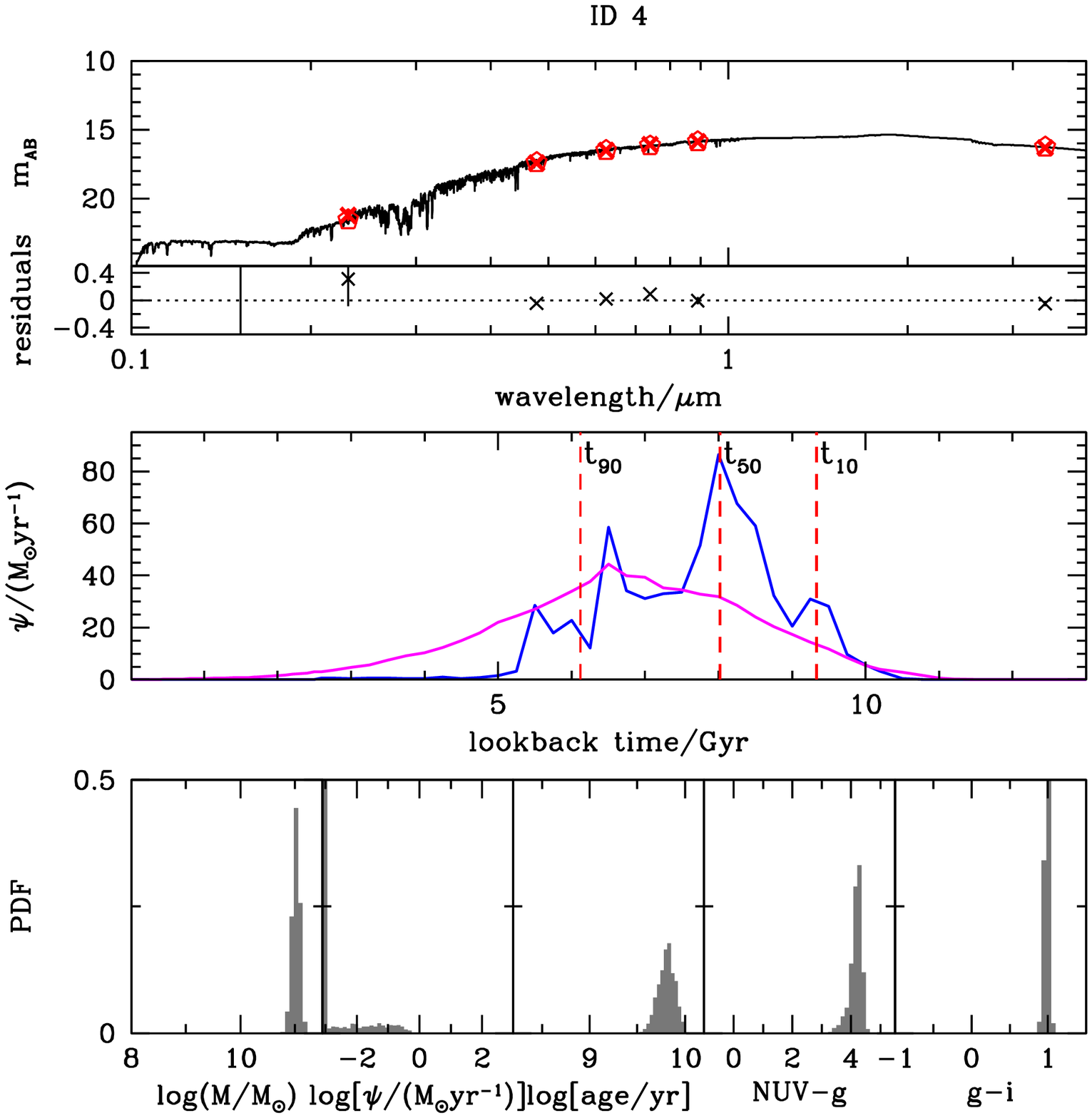}
\includegraphics[width=0.32\textwidth]{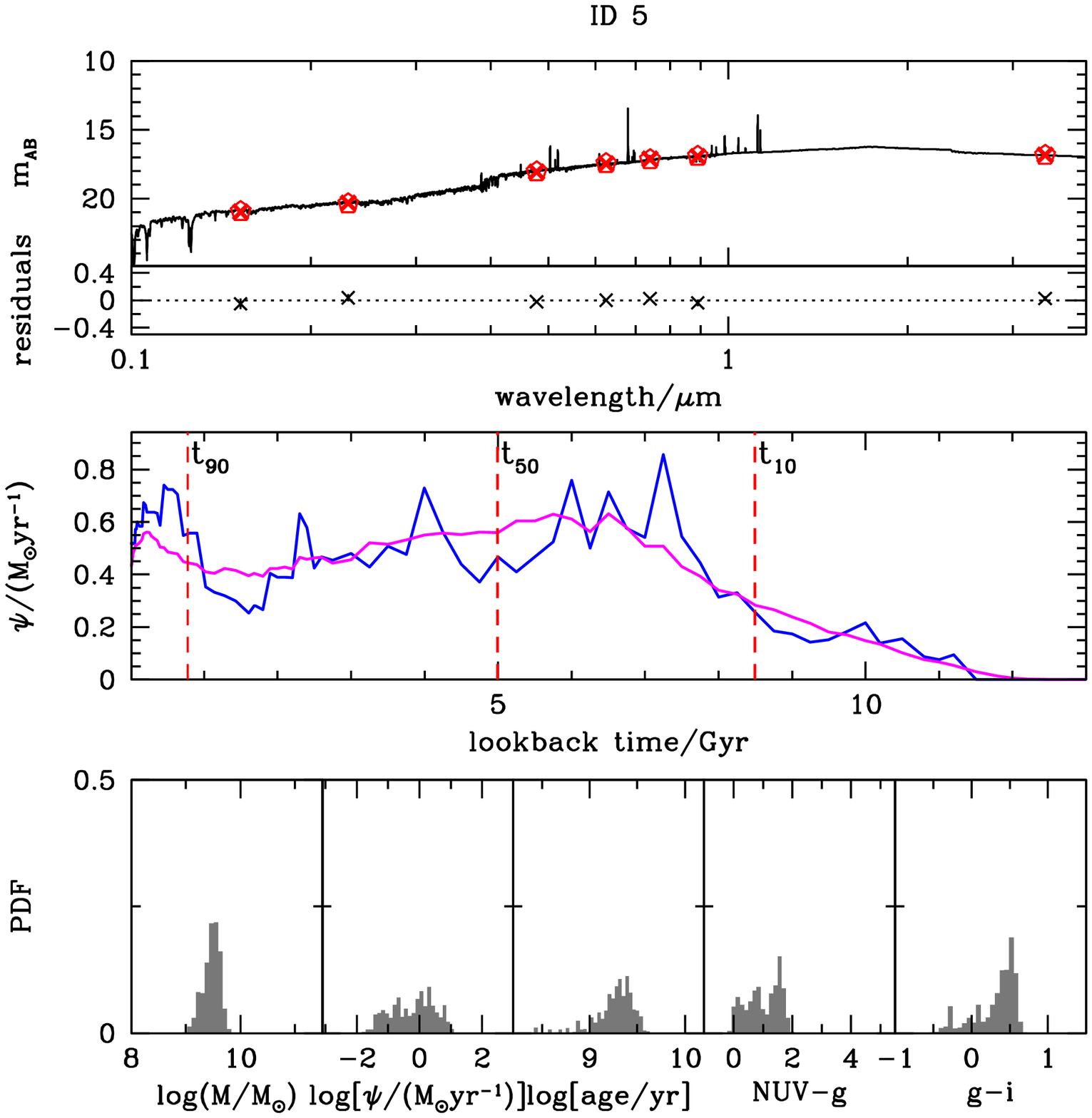}
\includegraphics[width=0.32\textwidth]{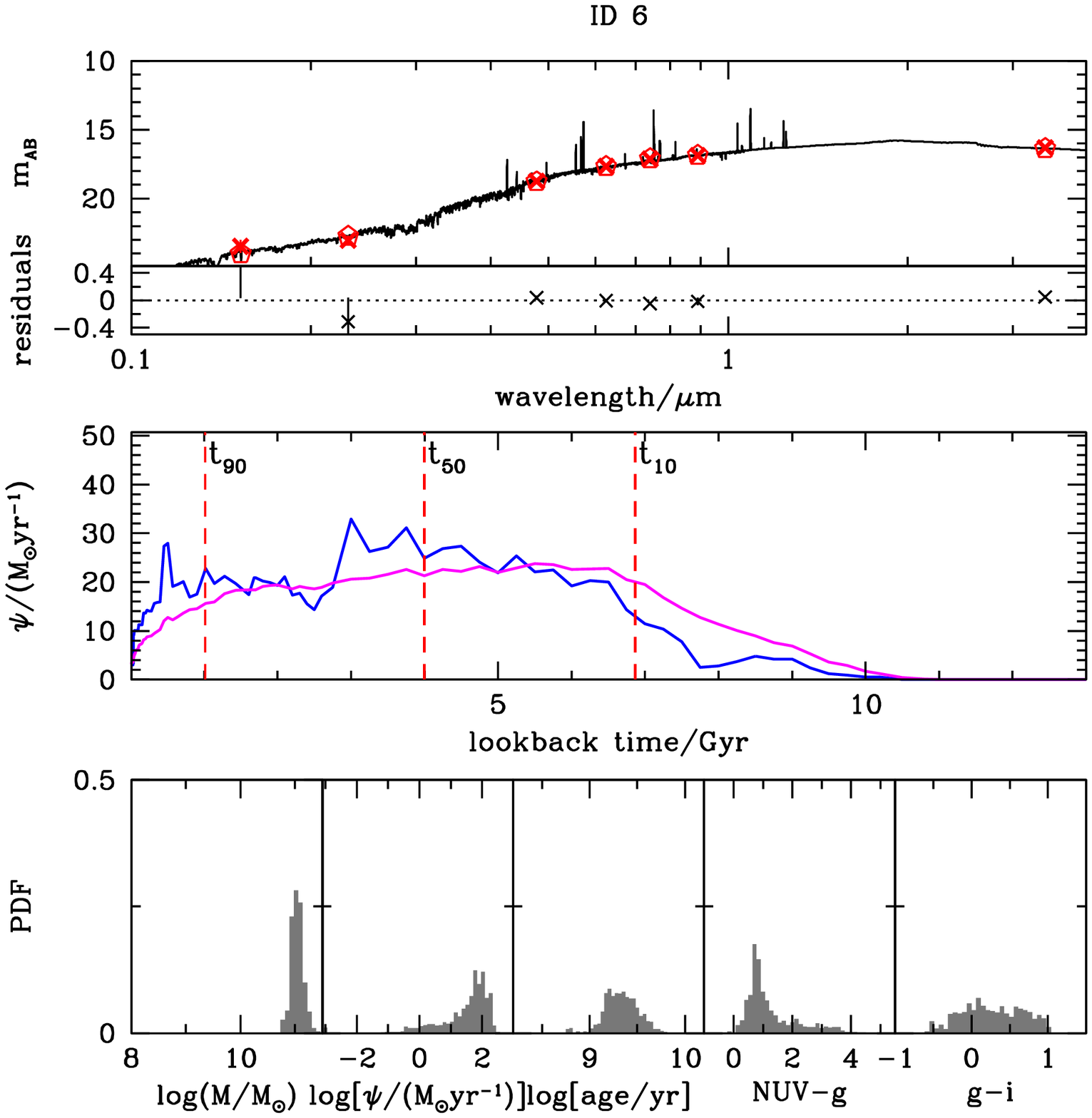}
\includegraphics[width=0.32\textwidth]{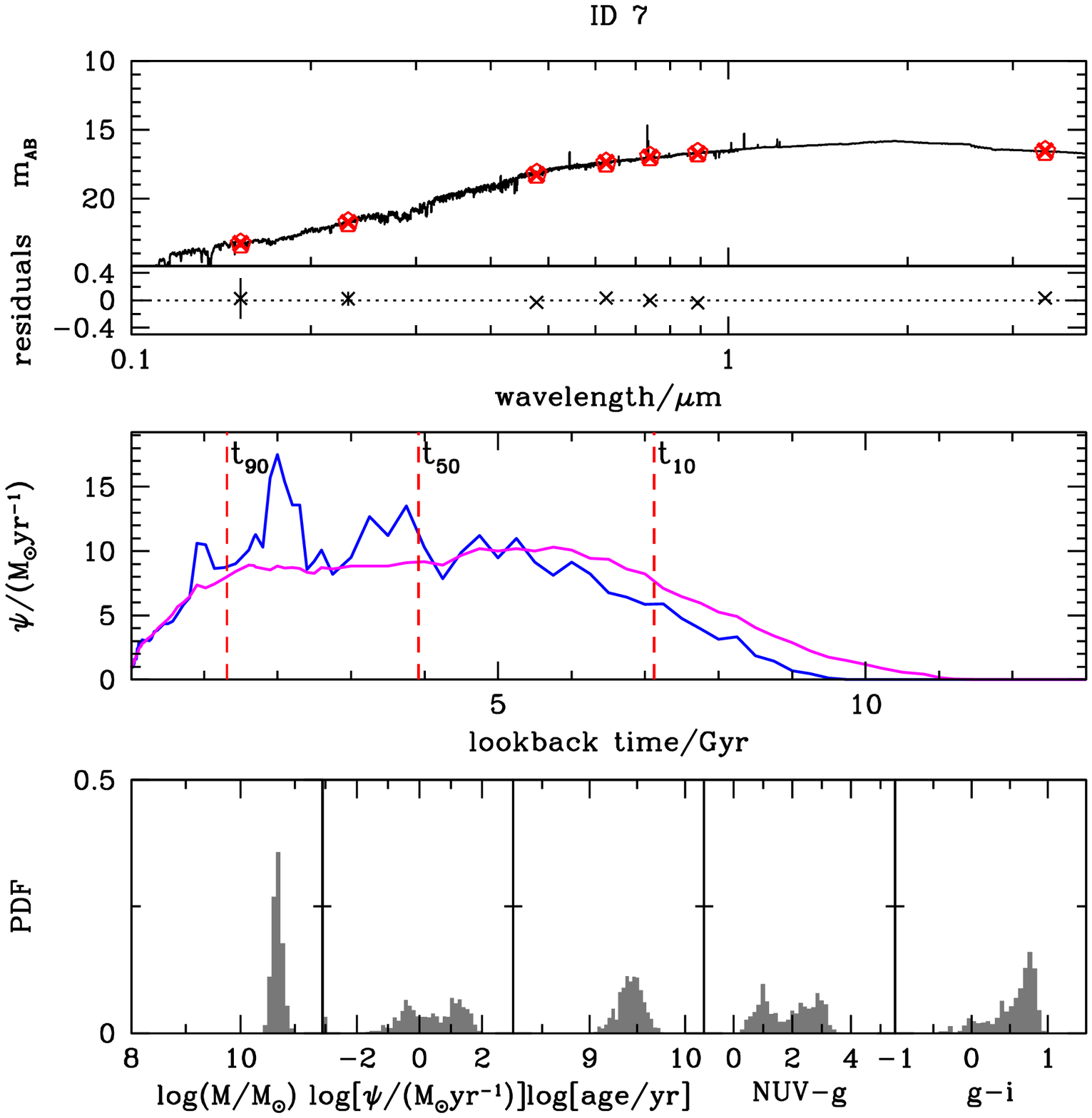}
\includegraphics[width=0.32\textwidth]{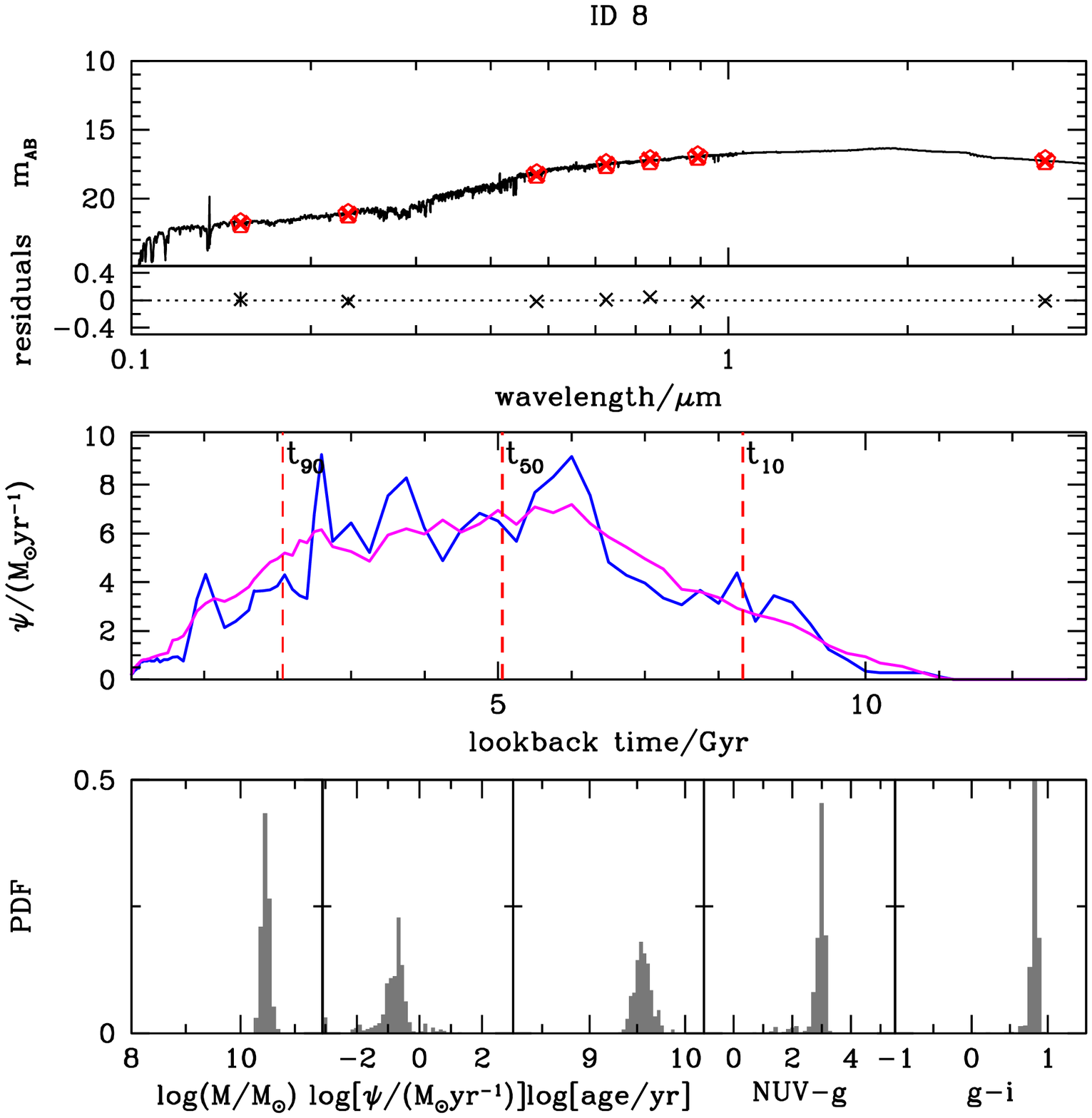}
\includegraphics[width=0.32\textwidth]{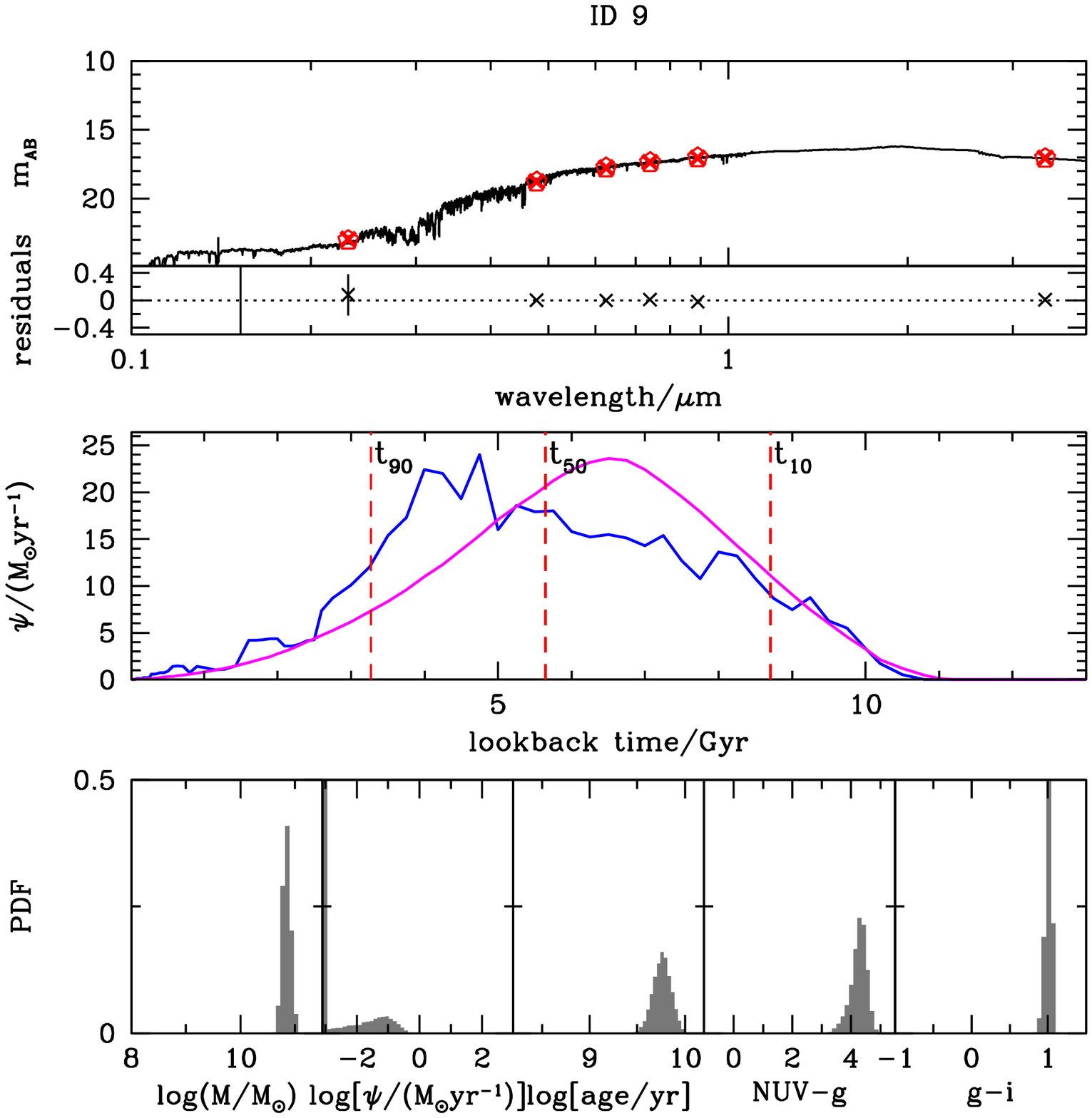}
\includegraphics[width=0.32\textwidth]{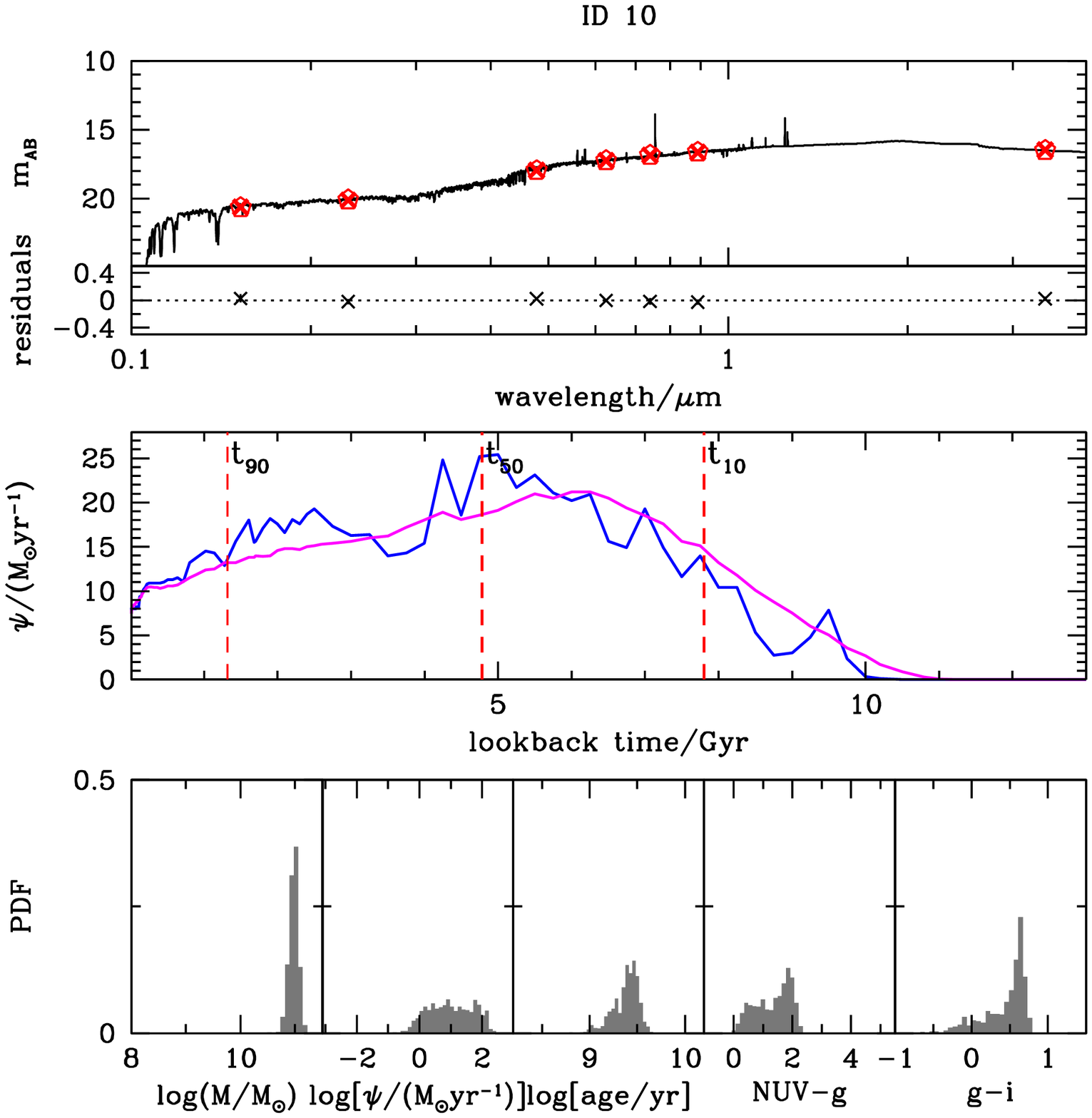}
\caption{Fits to different galaxies as shown in Figure~\ref{fig:fit}. Top panel: observed photometric magnitudes (red crosses), best-fit photometric magnitudes (red open diamonds), best-fit SED in full resolution (black solid line), and residuals between the observed and best-fit magnitudes (black crosses). Middle panel: likelihood-weighted average SFH derived from the first 10 best-fit models (blue solid line) and from all models (magenta solid line); lookback times at which the galaxy reaches 10\% ($t_{10}$), 50\% ($t_{50}$), and 90\% ($t_{90}$) of the total stellar mass formed (red vertical dashed lines) according to the blue SFH. Bottom panels: probability density function of (from left to right) stellar mass, SFR, light-weighted age, and rest-frame dust-corrected $NUV-g$ and $g-i$ colors.}
\label{fig:appendix}
\end{center}
\end{figure*}

\end{document}